\documentclass[conference]{IEEEtran}
\IEEEoverridecommandlockouts
\usepackage{cite}
\usepackage{amsmath,amssymb,amsfonts}
\usepackage{algorithmic}
\usepackage{graphicx}
\usepackage{textcomp}
\usepackage{xcolor}
\usepackage{multirow}
\usepackage{flushend}
\usepackage[scaled]{helvet}
\usepackage[T1]{fontenc}
\usepackage[left=0.625in, right=0.625in, top=0.75in,
bottom=1.02in]{geometry}
\usepackage[ruled, lined, longend, linesnumbered]{algorithm2e}
\begin{document}

\title{Resource Allocation in STAR-RIS-Aided SWIPT with RSMA via Meta-Learning}

\vspace{-0.50cm}
\author{\IEEEauthorblockN
	{Mojtaba~Amiri$^{\dag}$, Elaheh~Vaezpour$^{\S}$, Sepideh~Javadi$^{\star}$, Mohammad~Robat Mili$^{\star}$,\\ Halim~Yanikomeroglu$^{\star\star}$, and Mehdi~Bennis$^{\dag\dag}$
	}
	\vspace{0.50cm}
	\\$^{\dag}$ School of Electrical and Computer Engineering, University of Tehran, Tehran, Iran
	\\$^{\S}$ Dept. of Communication, ICT Research Institute, Tehran, Iran\\
	$^{\star}$ Pasargad Institute for Advanced Innovative Solutions (PIAIS), Tehran, Iran\\
	$^{\star\star}$ Dept. of Systems and Computer Engineering, Carleton University, Ottawa, ON, Canada\\
	$^{\dag\dag}$ Centre for Wireless Communications, University of Oulu, Oulu, Finland\\
}
\maketitle
\author{\IEEEauthorblockN{Mojtaba Amiri}
\IEEEauthorblockA{\textit{School of Electrical and Computer Engineering} \\
\textit{University of Tehran}\\
 Tehran, Iran \\
mojtaba.amiri@ut.ac.ir}
\and
\IEEEauthorblockN{Elaheh Vaezpour}
\IEEEauthorblockA{\textit{Communication Department} \\
\textit{ICT Research Institute}\\
Tehran, Iran \\
e.vaezpour@itrc.ac.ir}
\and
\IEEEauthorblockN{Sepideh Javadi}
\IEEEauthorblockA{\textit{dept. name of organization (of Aff.)} \\
\textit{Pasargad Institute for Advanced Innovative Solutions (PIAIS)}\\
Tehran, Iran\\
sepideh.javadi@piais.ir}
\and
\IEEEauthorblockN{Mohammad Robat Mili}
\IEEEauthorblockA{\textit{dept. name of organization (of Aff.)} \\
\textit{Pasargad Institute for Advanced Innovative Solutions (PIAIS)}\\
Tehran, Iran\\
mohammad.robatmili@gmail.com}
\and
\IEEEauthorblockN{Halim Yanikomeroglu}
\IEEEauthorblockA{\textit{Department of Systems and Computer Engineering} \\
\textit{Carleton University}\\
Ottawa, ON, Canada\\
halim@sce.carleton.ca}
\and
\IEEEauthorblockN{Mehdi Bennis}
\IEEEauthorblockA{\textit{Centre for Wireless Communications} \\
\textit{University of Oulu}\\
90570 Oulu, Finland \\
 mehdi.bennis@oulu.fi}
}

\maketitle

\begin{abstract}
		Simultaneously transmitting and reflecting reconfigurable intelligent surface (STAR-RIS) is a cutting-edge concept for the sixth-generation (6G) wireless networks. In this paper, we propose a novel system that incorporates STAR-RIS with simultaneous wireless information and power transfer (SWIPT) using rate splitting multiple access (RSMA). The proposed system facilitates  communication from a multi-antenna base station (BS) to single-antenna users in a downlink transmission. The BS concurrently sends energy and information signals to multiple energy harvesting receivers (EHRs) and information data receivers (IDRs) with the support of a deployed STAR-RIS. Furthermore, an optimization is introduced to strike a balance between users' sum rate and the total harvested energy. To achieve this, an optimization problem is formulated to optimize the energy/information beamforming vectors at the BS, the phase shifts at the STAR-RIS, and the common message rate.
		Subsequently, we employ a meta deep deterministic policy gradient (Meta-DDPG) approach to solve the complex problem. Simulation results validate that the proposed algorithm significantly enhances both data rate and harvested energy in comparison to conventional DDPG.
\end{abstract}


\section{Introduction}
	The concept of reconfigurable intelligent surface (RIS) is a key sixth-generation (6G) technology, attracting noteworthy attention \cite{basharat2021reconfigurable}. More specifically, each RIS consists of plenty of reflection units capable of adjusting amplitude and phase shift independently using a smart controller \cite{wang2023road, SIRS-NOMA}. Typically, RISs can only reflect the received incident signal and thus working in one half-plane. Therefore, a new advanced type of RIS, simultaneously transmitting and reflecting reconfigurable intelligent surface (STAR-RIS), is recommended to overcome this limitation by enabling the full-space coverage \cite{10@Ahmed,mu2024reconfigurable}. More specifically, the authors in \cite{5@Perera} consider the deployment of STAR-RIS in a full-duplex system aiming to maximize the average weighted sum rate.
	Moreover, a joint deployment and beamforming design algorithm is proposed in \cite{9@Pan} to address the sum rate maximization problem in a downlink STAR-RIS-assisted communication system.
	
	On the other hand, rate-splitting multiple access (RSMA) is a novel framework, offering substantial gains in spectral and energy efficiency\cite{mao2022rate, RSMA2, RSMA3, RSMA4, RSMA5}.
	Motivated by the advantages of the integration of RIS/STAR-RIS with RSMA, a few existing works have explored the performance of this integration \cite{1@Yang, 3@Meng, 7@Sun}. More specifically, the authors in \cite{1@Yang} focus on maximizing the energy efficiency (EE) in a downlink RIS-aided communication system with RSMA. In \cite{3@Meng},  a proximal policy optimization method is introduced to address the sum-rate maximization problem in a downlink STAR-RIS-assisted network with RSMA. Additionally, the authors in \cite{7@Sun} have explored a two-user RIS-aided RSMA system in the context of uplink transmission, with the goal of maximizing the achievable rate.
	
	Simultaneous wireless information and power transfer (SWIPT) is a cutting-edge technology addressing energy limitations in battery-powered devices \cite{goktas2023irs, J1, J2, SWIPTS}. Furthermore, optimizing both wireless communication performance and energy transfer is crucial, leading to optimization approaches \cite{khalili2020performance,mili2019novel}. More specifically, the sum rate maximization problem of multi-user SWIPT system with the deployment of an intelligent reflecting surface (IRS) is studied in \cite{zargari2021max}, while the authors in \cite{111Ren} aim to maximize the weighted sum rate of the information users in an active IRS-assisted SWIPT system. Moreover, the EE maximization problem of the active STAR-RIS-aided non-linear SWIPT system is investigated in \cite{SWIPTF}, where the authors have applied both mathematical and learning methods to solve the non-convex problem. The authors in \cite{SWIPTG} aim to maximize the EE of the IRS-aided SWIPT system, where an alternating optimization (AO) algorithm is employed to solve the problem.
	
	It should be noted that this is the first work investigating a multi-user STAR-RIS-assisted SWIPT system with RSMA. In this setup, users are divided into two main groups, i.e., information decoding receivers (IDRs) and energy harvesting receivers (EHRs), to receive the signals reflected by a deployed STAR-RIS. The system performance is evaluated by jointly maximizing the total sum rate of IDRs and harvested energy by EHRs. Additionally,  different from the conventional DDPG, a Meta-learning-based DDPG algorithm is formulated to solve the non-convex NP-hard problem.
	\\
	\begin{figure}
		\centering
		\includegraphics[width=9cm, height=6cm]{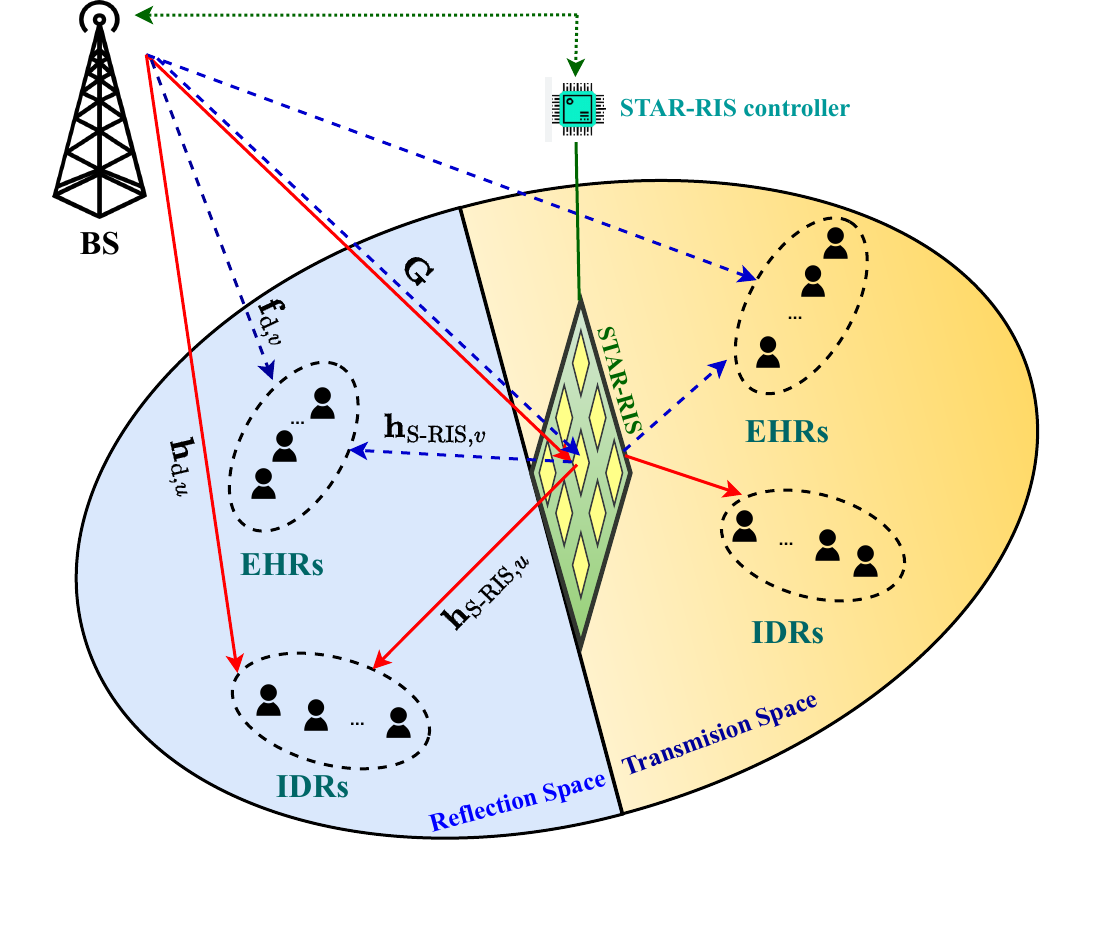}
		\caption{\footnotesize{System model of an STAR-RIS SWIPT system with RSMA.}}
		\label{Fig1}
	\end{figure}
The remainder of this paper is organized as follows. In Section II, the system model of a downlink STAR-RIS-assisted multiuser SWIPT system with RSMA is introduced. Section III formulates the optimization problem. In order to solve this problem, the Meta-DDPG algorithm is employed, detailed in Section IV. Furthermore, simulation results are expressed in Section V, while complexity analysis is shown in Section VI. Finally, Section VII concludes the paper.

\emph{Notation}: In this paper, boldface capital letters and lowercase letters represent matrices and vectors,
 respectively. $\mathbb{E}(.)$ and $(.)^{H}$ denote the expectation value and conjugate transposition, respectively, while $\mathbb{C}$ denotes the complex matrix.
\section{System Model}
	In this section, we consider a downlink STAR-RIS-assisted multi-user SWIPT system with RSMA, where an STAR-RIS composed of $M$ elements is implemented to assist the transmission between an $N$-antenna base station (BS) and $K$ single-antenna mobile users. The $K$ receivers are separated into two groups, i.e., $U$ IDRs and $V$ EHRs, such that $U \cup V = K$ and $U \cap V = \emptyset$. The sets of RIS elements, BS antennas, IDRs and EHRs are represented by $\mathcal{M} = \left\{ {1,2,...,M} \right\}$, $\mathcal{N} = \left\{ {1,2,...,N} \right\}$, $\mathcal{U} = \left\{ {1,2,...,U} \right\}$, and $\mathcal{V} = \left\{ {1,2,...,V} \right\}$, respectively.
	The signal is divided into the refracted and reflected components by each element of a configurable and programmable STAR-RIS. The transmitted and reflected signals by the $m$-th element of STAR-RIS can be respectively described as
	\begin{equation}
		s^t_{m} = \sqrt{\beta^{t}_{m}}e^{j\theta^t_{m}}s_m, ~~\textrm{and}~~
		s^r_{m} = \sqrt{\beta^{r}_{m}}e^{j\theta^r_{m}}s_m,
	\end{equation}
	where $s_m$ denotes the signal incident on the $m$-th element of STAR-RIS and $j$ represents the imaginary unit. Furthermore, the amplitude and phase shift of the transmission and reflection coefficients of the $m$-th element of the deployed STAR-RIS are respectively denoted by $\{\sqrt{\beta^{t}_{m}}, \sqrt{\beta^{r}_{m}}\}$ and $\{\theta^t_{m},\theta^r_{m}\}$. It is worth pointing out that the amplitude and phase shift of $m$-th element of STAR-RIS in both reflection and refraction sides are within $[0,2\pi]$ and $[0,1]$, respectively. It is assumed that the reflection and transmission phase shifts $\{\theta^t_{m},\theta^r_{m}\}$ are independent, while the amplitude coefficients for both reflection and transmission are coupled. Moreover, the condition  $\beta^{r}_{m}+\beta^{t}_{m}=1$ should be satisfied, as the incident signal's energy should be equal to aggregate energy of reflected and transmitted signals, i.e., ${\left| {s_m^t} \right|^2} + {\left| {s_m^r} \right|^2} = {\left| {{s_m}} \right|^2}$.
	
	With the one layer RSMA scheme at the BS, the total message of each IDR $u \in \mathcal{U}$ is divided into two main parts, i.e., a common part ${\bold{w}_c}$ and a private part $\mathbf{w}_{k}$. Then, the total transmit signal at the BS by considering both IDRs and EHRs is given by
	\begin{equation}
		\mathbf{x} = {{\bold{w}_c}}{s^{\text{ID}}_c}+\sum_{u=1}^{U}\mathbf{w}_{u}s^{\text{ID}}_{u} + \sum_{v=1}^{V}\mathbf{p}_{v}s^{\text{EH}}_{v},
	\end{equation}
	where ${s^{\text{ID}}_c}$, ${s^{\text{ID}}_u}$, and ${s^{\text{EH}}_v}\in \mathbb{C}$ are the common information signal, the private information signal of IDR $u \in \mathcal{U}$, and the energy signal of EHR $v \in \mathcal{V}$ in the proposed system model. It is assumed that both common and private information signals are independent and identically distributed (i.i.d), i.e., $\mathbb{E}\{|s^{\text{ID}}_{c}|^2\}=1$ and $\mathbb{E}\{|s^{\text{ID}}_{u}|^2\}=1$, while the energy signals are independently generated from an arbitrary distribution with $\mathbb{E}\{|s^{\text{EH}}_{v}|^2\}=1$. Moreover, $\mathbf{w}_{c}\in \mathbb{C}^{N \times 1}$ and $\mathbf{w}_{u}\in \mathbb{C}^{N \times 1}$ denote the common and private information beam for the $u$-th IDR, while $\mathbf{p}_{v}\in \mathbb{C}^{N \times 1}$ represents the transmit energy beam for the $v$-th EHR.
	%\subsection{Channel Model}
	We presume that the links of channel experience quasi-static flat-fading, and the channel state information is accessible at both BS and the deployed STAR-RIS. Subsequently, the received signal at IDR $u$ can be formulated as
	
	\begin{equation}
		{y}^{\text{ID}}_{u} = \mathbf{h}^{H}_{u}\mathbf{x} + z^{\text{ID}}_{u},
	\end{equation}
	where ${z^{\text{ID}}_{u}} \sim \mathcal{C}\mathcal{N}\left( {0,\sigma _u^2} \right)$ is the white Gaussian noise of the $u$-th IDR with zero mean and variance $\sigma _u^2$. The equivalent channel from the BS to the STAR-RIS, the STAR-RIS to the $u$-th IDR, as well as the direct path from the BS to IDR $u$ is given by
	\begin{equation}
		\mathbf{h}^{H}_{u} \triangleq \mathbf{h}^{H}_{d,u} + \mathbf{h}^{H}_{\text{S-RIS},u}\boldsymbol{\Theta}^l\mathbf{G},
	\end{equation}
	where $\mathbf{G} \in \mathbb{C}^{M \times N}$, $\mathbf{h}_{d,u} \in \mathbb{C}^{N \times 1}$, and $\mathbf{h}_{\text{S-RIS},u} \in \mathbb{C}^{M \times 1}$  denote the channel matrix from the BS to the STAR-RIS, the channel vector from the BS to IDR $u$, and the channel vector from the STAR-RIS to the $u$-th IDR, respectively. Furthermore, the transmission ($l=t$) or reflection ($l=r$) beamforming vector is denoted by ${\boldsymbol{\phi}^l} = \left[ {\sqrt {\beta _1^l} {e^{j\theta _1^l}},\sqrt {\beta _2^l} {e^{j\theta _2^l}},...,\sqrt {\beta _M^l} {e^{j\theta _M^l}}} \right]$, in which ${\boldsymbol{\Theta}^l} = \text{diag}({\phi ^l})$, $l \in \left\{ {t,r} \right\}$ represents the corresponding diagonal beamforming matrix of the STAR-RIS in transmission/reflection region.
	It is assumed that the number of $V$ out of $K$ receivers can harvest energy. Subsequently, the received signal at the EHR $v\in \mathcal{V}$ can be written as
	\begin{equation}
		{y}^{\text{EH}}_{v}= \mathbf{f}^{H}_{v}\mathbf{x} + z^{\text{EH}}_{v},
	\end{equation}
	where ${z^{\text{EH}}_{v}} \sim \mathcal{C}\mathcal{N}\left( {0,\sigma _v^2} \right)$ is the AWGN of the $v$-th EHR with zero mean and variance $\sigma _v^2$. Additionally, the combined channel of the BS-(STAR-RIS)-user link for the $v$-th EHR is given by
	\begin{equation}
		\mathbf{f}^{H}_{v}= \mathbf{f}^{H}_{d,v}+ \mathbf{f}^{H}_{\text{S-RIS},v}\boldsymbol{\Theta}^l\mathbf{G},
	\end{equation}
	where $\mathbf{f}_{d,v}\in \mathbb{C}^{N \times 1}$ is the channel vector between the BS and $v$-th EHR, while   $\mathbf{f}_{\text{S-RIS},v}\in \mathbb{C}^{M \times 1}$ denotes the channel vector between the STAR-RIS and EHR $v$.
	
	Following this, each IDR decodes its private stream after eliminating the common stream using successive interference cancellation (SIC). For the sake of simplicity, it is assumed that all IDRs can mitigate interference caused by the energy signals \cite{zargari2021max}. Hence, the received information rates of the common stream and the private stream at the $u$-th IDR can be respectively stated as follows \cite{RSMA4, RSMA5}:
	\begin{equation}
		R_{c,u}={\log _2}\left( 1+ {\frac{|\mathbf{h}^{H}_{u}\mathbf{w}_{c}|^2}{\sum_{i=1}^{U}|\mathbf{h}^{H}_{u}\mathbf{w}_{i}|^2+\sigma_{u}^{2}}} \right),
	\end{equation}
	and
	\begin{equation}
		R_{p,u}={\log _2}\left(1+{\frac{|\mathbf{h}^{H}_{u}\mathbf{w}_{u}|^2}{\sum_{i=1,i\neq u}^{U}|\mathbf{h}^{H}_{u}\mathbf{w}_{i}|^2+\sigma_{u}^{2}}} \right).
	\end{equation}
	To guarantee the successful decoding of the common message at all IDRs, the achievable rate of the common message should not exceed the minimum rate of all IDRs, i.e., $R_{c}\leq \min \lbrace R_{c,1}, R_{c,2}, \dots, R_{c,U} \rbrace$. The common rate $R_{c}=\sum_{u=1}^{U}C_{u}$ is shared by all $U$ IDRs, while $C_{u}$ denotes the $u$-th IDR's portion of common rate. Then, we have
	\begin{equation}
		\sum_{u=1}^{U} C_{u} \leq \min_u \lbrace R_{c,u}\rbrace, \forall u \in \mathcal{U}.
	\end{equation}
	Furthermore, the total harvested energy at the $v$-th EHR is expressed as \cite{wu2019weighted}
	\begin{equation}
		P^\text{Har}_{v}(\mathbf{w}_{u},\mathbf{p}_{v},\bold{\Theta}^l) = \lambda_{v}\mathbb{E}\{\sum_{u=1}^{U}|\mathbf{f}^{H}_{v}\mathbf{w}_{u}|^2+\sum_{v=1}^{V}|\mathbf{f}^{H}_{v}\mathbf{p}_{v}|^2\},
	\end{equation}
	where $\lambda_{v}\in [0,1]$ represents the energy conversion efficiency of the EHR $v$.
	\section{Optimization Problem Formulation}
	In this paper, we specifically address a maximization problem for optimizing both the sum rate of IDRs and the total harvested energy at the EHRs by jointly optimizing the active beamforming vectors at the BS, i.e., $\mathbf{W} = \lbrace \mathbf{w}_{c},\mathbf{w}_{u}, \forall u \rbrace$ and $\mathbf{P} = \lbrace \mathbf{p}_{v}, \forall v \rbrace$, data rate for receiving the common message $\mathbf{c} = \lbrace C_{u}, \forall u  \rbrace$, and transmission/reflection coefficients of STAR-RIS. The total sum-rate and the total harvested energy can be normalized as $R_{\text{Norm}} = \frac{\sum_{u=1}^{U}{(R_{c,u}+R_{p,u})}}{\text{max}{\{R_{c,u}+R_{p,u}\}}}$ and $ P^{\textrm{Har}}_{\text{Norm}} = \frac{\sum_{v=1}^{V} P^{\textrm{Har}}_{v}}{\text{max}\{P^{\textrm{Har}}_{1},...,P^{\textrm{Har}}_{V}\}}$. Then, the maximization problem is given by
	\begin{alignat}{5}\nonumber
		\max_{\mathbf{W}, \mathbf{P}, \boldsymbol{\Theta}^l, \mathbf{c}} \quad &\alpha R_{\text{Norm}} + (1-\alpha) P^{\textrm{Har}}_{\text{Norm}} ,\\\nonumber
		%\max_{\mathbf{W}, \mathbf{P}, \boldsymbol{\Theta}^l} \quad & \sum_{v=1}^{V}P^\text{Har}_{v},\\\nonumber
%		\textrm{s.t.}
		\quad & \text{C}_1: \sum_{u=1}^{U} C_{u} \leq \min_u \lbrace R_{c,u}\rbrace, \quad \forall u \in \mathcal{U},  \\\nonumber
		\quad & \text{C}_2: C_{u}+R_{u}\geq R_{\text{min},u}, \quad  \forall u \in \mathcal{U},    \\\nonumber
		\quad & \text{C}_3: {\parallel \mathbf{w}_{c} \parallel^2} +\sum_{u=1}^{U}{\parallel \mathbf{w}_{u} \parallel^2} + \sum_{v=1}^{V}\parallel \mathbf{p}_{v} \parallel^2 \leq P_{\text{max}},\\\nonumber
		\quad & \text{C}_4: P^\text{Har}_{v} \geq E_{\text{min},v}, \quad  \forall v \in \mathcal{V},   \\\nonumber
		\quad & \text{C}_5: { {{\beta _{m}^t} ,{\beta _{m}^r}  \in \left[ {0,1} \right]}}, \quad \beta_{m}^t + \beta _{m}^r = 1, ~ \forall m \in \mathcal{M}, \\
		\quad & \text{C}_6: {\theta _{m}^t,\theta_{m}^r \in \left[ {0,2\pi } \right)}, \quad \forall m \in \mathcal{M}, \label{eq:constraint1}
	\end{alignat}
	where $\alpha \in [0,1]$ is a weighted factor. Constraints $\text{C}_1$ and $\text{C}_2$ ensure that the common message
	can be successfully decoded at all the IDRs in both reflection and transmission spaces. Moreover, constraint $\text{C}_3$ denotes the maximum power budget at the BS, while $\text{C}_4$ represents the minimum harvested energy for the $v$-th EHR.
	Finally, the constraints for amplitude and phase shift coefficients of the deployed STAR-RIS are denoted by $\text{C}_5$ and $\text{C}_6$, respectively. The above non-convex optimization problem is NP-hard, making it challenging to solve using conventional solutions. Therefore, a Meta-DDPG approach is applied to solve this problem.	
\section{Meta-DDPG based Learning Solution}
Previous research using convex optimization techniques suggests intricate mathematical transformations for a local optimal solution. Furthermore, given the dynamic nature of wireless environments, real-time adaptability is preferred. Recent DRL methods achieve this and propose a unified solution without decomposing the problem, unlike prior convex optimization approaches. Therefore, in this paper, a Meta-DDPG approach is employed to address the optimization issue.
	The Meta-DDPG approach \cite{hospedales2021meta} is a reinforcement learning (RL) method employed due to the extensive dimensionality of both the state and action spaces.
	Conventional DDPG algorithm cannot quickly adapt to new and dynamic environments. Instead, Meta-learning is used to address this issue. This algorithm integrates conventional DDPG with Meta-learning to enable learning a model which rapidly adapts to a new environment. This method is based on the agent, state, action, and reward \cite{luong2019applications, hospedales2021meta,maleki2022multi} and utilizes actor and critic networks to learn a suitable model.
	
	In each time step $t \in [0,T]$, the agent observes a candidate state of the environment represented as $s_{t} \in \mathcal{S}$, and subsequently selects an action based on the current state $s_{t}$ which is denoted by $a_{t} \in \mathcal{A}$. After taking action $a_{t}$ in state $s_{t}$ by the agent, the environment transits to a new state $s_{t+1} \in \mathcal{S}$, in which the agent receives a numerical reward $r_{t+1} \in \mathcal{R}$. As a result, a sequence of interactions,  i.e., $s_0, a_0, r_1, s_1, a_1, r_2, \ldots$,  unfolds over time. The value of $r$ plays a crucial role as it guides the agent in selecting better actions. The policy, denoted as $\pi$, defines how the agent maps states to the actions.
	%In a deterministic policy, a unique action is specified for each state, ensuring that the agent consistently chooses the same action in a given state.
	In the Meta-DDPG algorithm, the main purpose is to find an optimal policy denoted as $\pi^*$, with the ultimate aim of maximizing the reward function. To characterize the Meta-DDPG algorithm, we examine its components as follows:\\
	1) State: It is defined as a set of channel information, ensuring that the states have the necessary information to make proper decisions.\\
	2) Action: At each time step $t$, the agent selects the action $ a_{t} = [\mathbf{W},\mathbf{P},\boldsymbol{\Theta}^l, \mathbf{c}]$ based on the current state $s_t$.\\
	3) Reward function: As the reward is an important part of RL, it should be designed properly to ensure the learning model, aimed at solving the optimization problem, satisfies the constraints. Therefore, the reward is defined as follows:
	\begin{eqnarray}
		r_t=\left\lbrace \begin{array}{lc}
			\Xi, & \text{if constraints are satisfied},\\
			0, & \text{otherwise},\\
		\end{array}\right.
	\end{eqnarray}
	where $\Xi = R_{\text{Norm}} + (1-\alpha) P^{\textrm{Har}}_{\text{Norm}}$ and a penalty with the value of $0$ is imposed to discourage the selection of actions that do not satisfy the constraints.
	%\vspace{-0.5cm}
	\begin{algorithm}[b!]
		\caption{Meta-DDPG algorithm}
		\label{algorithm1}
		\textbf{Input}: The maximum number of time steps (T), the number of episodes (E), $\xi$ and $X$. \\
		Initialize the actor and critic networks, $\pi(s|\theta^\pi)$ and $Q(s,a|\theta^Q)$.\\
		Initialize the target critic and actor networks with parameters of
		${\bar{\theta}}^Q \leftarrow \theta^Q$ and $\bar{\theta}^\pi \leftarrow \theta^\pi$.\\
		Initialize a replay buffer $B$.
		\\
		\For{$episode=1$ \KwTo $E$}{
			initialize observation state $s_0$
			\\
			\For{$t=1$ \KwTo $T$}{
				Choose action $a_t$ based on the current policy and exploration noise.\\
				Execute action $a_t$, observe reward $r_t$ and record the new state $s_{t+1}$ by storing the transition $(s_{t}, a_{t}, r_{t}, s_{t+1})$ in $ B$.
				\\
				\For{each step of gradient descent to solve problem \text{(\ref{eq:constraint1})}}{
					A mini-batch is randomly selected from the set $B$.
					\\
					$\theta^Q = \theta^Q - \text{lr}_{\text{critic}}\nabla L(\theta^Q)$
					\\
					$\theta^\pi_{\text{old}} = \theta^\pi - \text{lr}_{\text{actor}}\nabla J(\theta^\pi) $
					\\
					$\theta^\pi_{\text{new}} = \theta^\pi_{\text{old}}- \text{lr}_{\text{actor}}\nabla F_{\text{new}}(\theta^\pi,\mu)$
					\\
					$\theta^\pi \leftarrow \theta^\pi_{\text{new}}$
					\\
					$\mu = \mu - \text{lr}_{\text{meta}}\nabla \text{tanh}(J(\theta^\pi_{\text{new}})-J(\theta^\pi_{\text{old}}))$
					\\
					\If{$\text{mod}(t,X) = 0$}{
						${{\bar{\theta}}^Q}_{t+1} = (1-\xi){\bar{\theta}^Q}_{t} + {\xi{\theta_{t}^Q}}$
						\\
						${{\bar{\theta}}^\pi}_{t+1} =(1-\xi){{\bar{\theta}}^\pi}_{t} + \xi{\theta_{t}^\pi}$
					}}}}
				\end{algorithm}
				
				Here, the Meta-DDPG solution is proposed, merging traditional DDPG with Meta-learning for swift adaptation to new environments. Actually, Meta-DDPG extends traditional DDPG by adding Meta-learning. It has inner loop for fine-tuning and outer loop for faster adaptation to new environments. Note that the parameters of four neural networks are represented as follows: critic network $\theta^Q$, actor network $\theta^\pi$, target critic network ${\bar{\theta}}^Q$, and target actor network $\bar{\theta}^\pi$.
				\\The actor network parameters are adjusted in a manner that minimizes the subsequent loss function:
				\begin{equation}
					J(\theta^\pi) = -Q^{\pi}(s_{t},a_{t}=\pi(s|\theta^\pi)|\theta^Q),
				\end{equation}
				where $\pi(s|\theta^\pi)$ and $Q(s,a|\theta^Q)$ are actor and critic networks, respectively. In order to learn the parameters of critic network, the subsequent loss function is minimized:
				\begin{equation}
					L(\theta^Q) = \mathbb{E}[(Q(s_{t},a_{t}|\theta^Q)-Y)^2],
				\end{equation}
				where $Y = R + \tau Q(s_{t+1},\pi(s_{t+1}|\theta^\pi)|\theta^Q)$ and $R = \sum_{t=0}^{T}\tau^{t}r_{t}$ with $\tau \in [0,1)$. Subsequently, to update the parameters of the target critic and target actor networks, ${{\bar{\theta}}^Q}_{t+1} = (1-\xi){\bar{\theta}^Q}_{t} + {\xi{\theta_{t}^Q}}$ and
				${{\bar{\theta}}^\pi}_{t+1} =(1-\xi){{\bar{\theta}}^\pi}_{t} + \xi{\theta_{t}^\pi}$ with $\xi\ll1$ are defined.
				
				A tuple $(s_t, a_t, r_t,s_{t+1})$ is stored into a buffer pool and then is updated within different episodes. The bi-level optimization problem is defined as follows:
				\begin{subequations}
					\begin{equation}
						\mu=\text{arg}~\underset{\mathbf{\mu}}{\min}\ F_{\text{meta}}(\theta^{*\pi}),
					\end{equation}
					subject to
					\begin{flalign}
						& \theta^{*\pi} = \text{arg}~\min_{\theta^\pi}(J(\theta^\pi)+F_{\text{new}}(\theta^\pi,\mu))
						,\label{eq:c43}
					\end{flalign}
				\end{subequations}
				where $F_{\text{meta}}(\theta^{*\pi}) = \text{tanh}(J(\theta^\pi_{\text{new}})-J(\theta^\pi_{\text{old}}))$ and $F_{\text{new}}(\theta^\pi,\mu)=\mathbb{E}[\mu log(1+e^{\pi(s|\theta^\pi)}]$.
				In the context of the DDPG algorithm, the actor parameters are updated according to the equation $\theta^\pi_{old} = \theta^\pi - \text{lr}_{\text{actor}}\nabla J(\theta^\pi) $. Furthermore, in the Meta-DDPG variant, the actor parameters are derived from a distinct process denoted as $\theta^\pi_{\text{new}} = \theta^\pi_{\text{old}}- \text{lr}_{\text{actor}}\nabla F_{\text{new}}(\theta^\pi,\mu)$.
				The incorporation of these additional steps yields superior performance and enhanced convergence compared to the conventional DDPG algorithm.
				\section{Simulation Results}
				The effectiveness of the proposed method is empirically assessed with a comparison to classical methods. In the simulations, we set $M=20$, while $K$ is equal to 4. Additionally, it is supposed that the BS is equipped with $N = 4$ antennas, positioned at coordinates (0,0)~m. Meanwhile, the RIS is located at (100,0)~m. Furthermore, the system operates within defined constraints, including $P_{\text{max}} = 20~\textrm{dBm}$, $E_{\text{min}} = 0.0004$, and $R_{\text{min}}= 0.1$ bps/Hz.
				
\begin{figure}
					\centering
					\includegraphics[width=9cm, height=6cm]{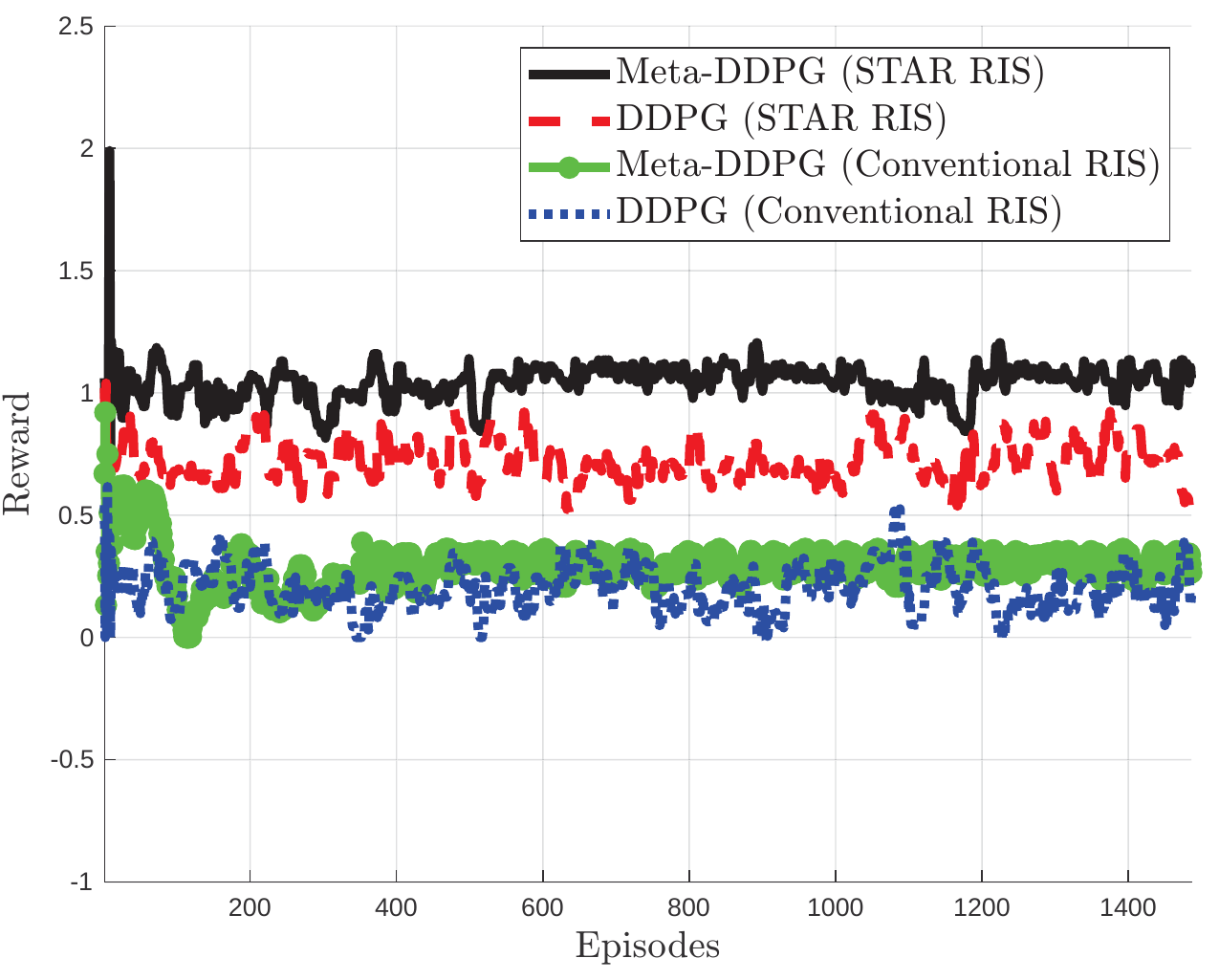}
					\caption{\footnotesize{Convergence of the Meta-DDPG-based algorithm with $\alpha=0.5$.}}
					\label{convergence}
				\end{figure}

				In Fig. \ref{convergence}, the reward for various episodes is shown, where the Meta-DDPG depicts faster and smoother convergence compared to the DDPG approach. Moreover, combining the Meta-DDPG with STAR-RIS results in even faster convergence than with conventional RIS, highlighting the significant improvement in convergence speed. It is noteworthy that the Meta-DDPG with STAR-RIS outperforms the DDPG with STAR-RIS, emphasizing the advantages of Meta-learning with STAR-RIS.
				
\begin{figure}
					\centering
					\includegraphics[width=9cm, height=6cm]{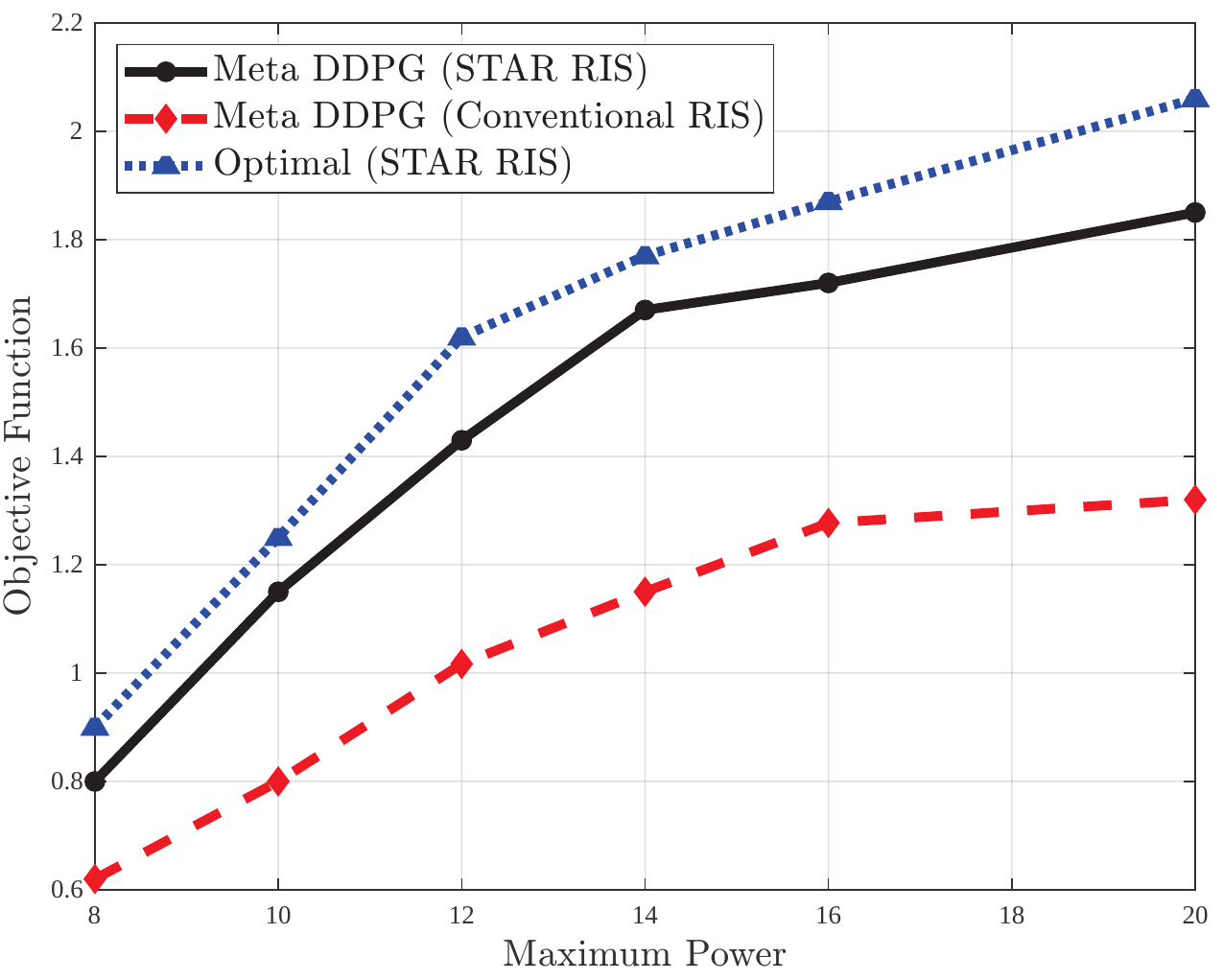}
					\caption{\footnotesize{The objective versus maximum transmit power with $\alpha=0.5$.}}
					\label{MP}
				\end{figure}

				In Fig. \ref{MP}, the objective function versus maximum transmit power in the BS for $\alpha=0.5$ is shown.
				We observe that the objective improves for different methods by increasing the maximum transmit power. The maximum transmit power of the BS is varied across a range of values, spanning from 8 dBm to $P_{\text{max}}$. The objective function for the STAR-RIS scenario consistently outperformed that of the conventional RIS scenario. Notably, the performance of this approach is close to the optimal scenario, where the exhaustive search method is considered as the optimal approach. This indicates that STAR-RIS demonstrates superior performance compared to conventional RIS, and it is remarkably competitive with the theoretically optimal scenario.
				
\begin{figure}
					\centering
					\includegraphics[width=9cm, height=6cm]{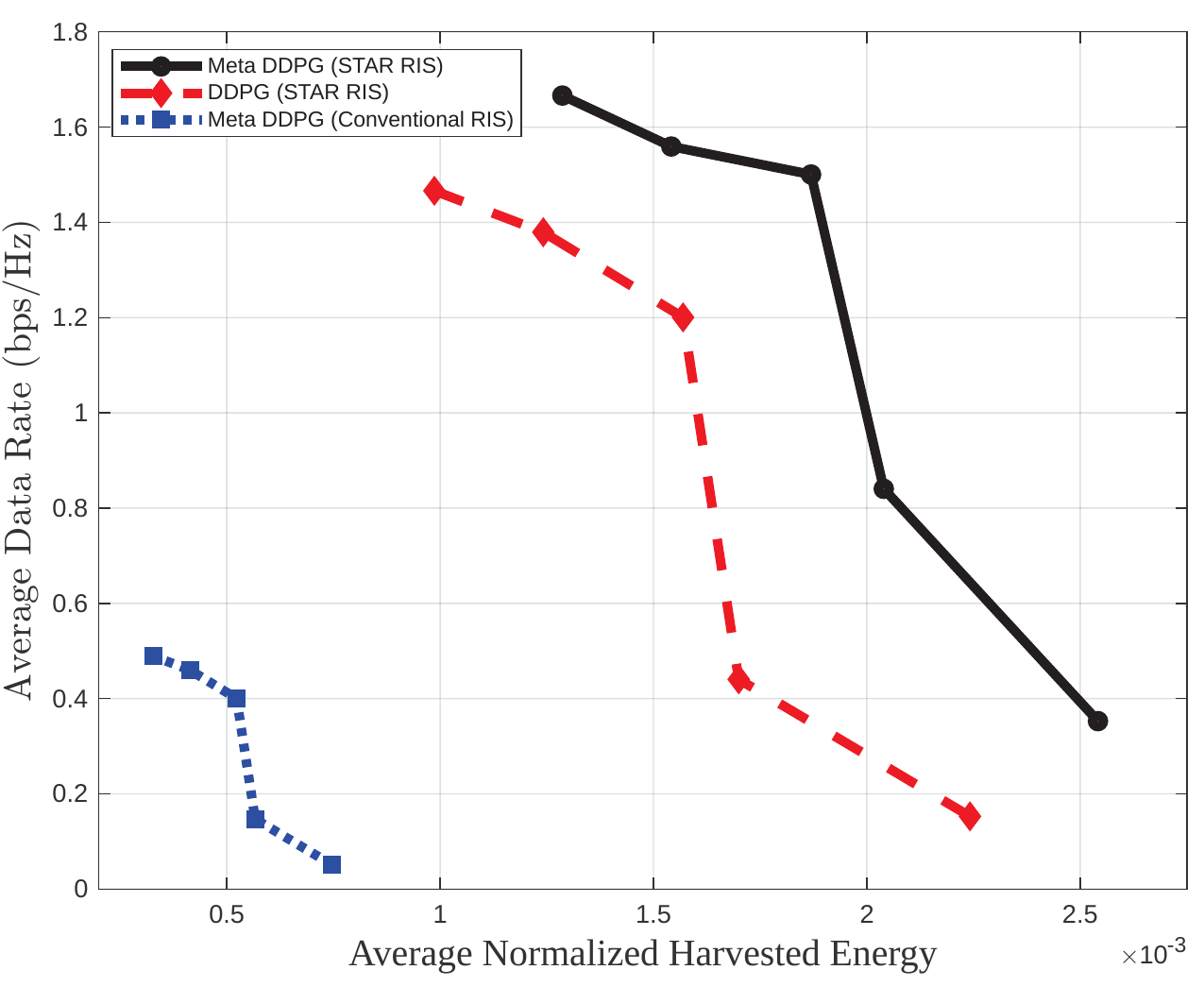}
					\caption{\footnotesize{Data rate versus ANHE.}}
					\label{EHADR}
				\end{figure}

\begin{figure}
	\centering
	\includegraphics[width=9cm, height=6cm]{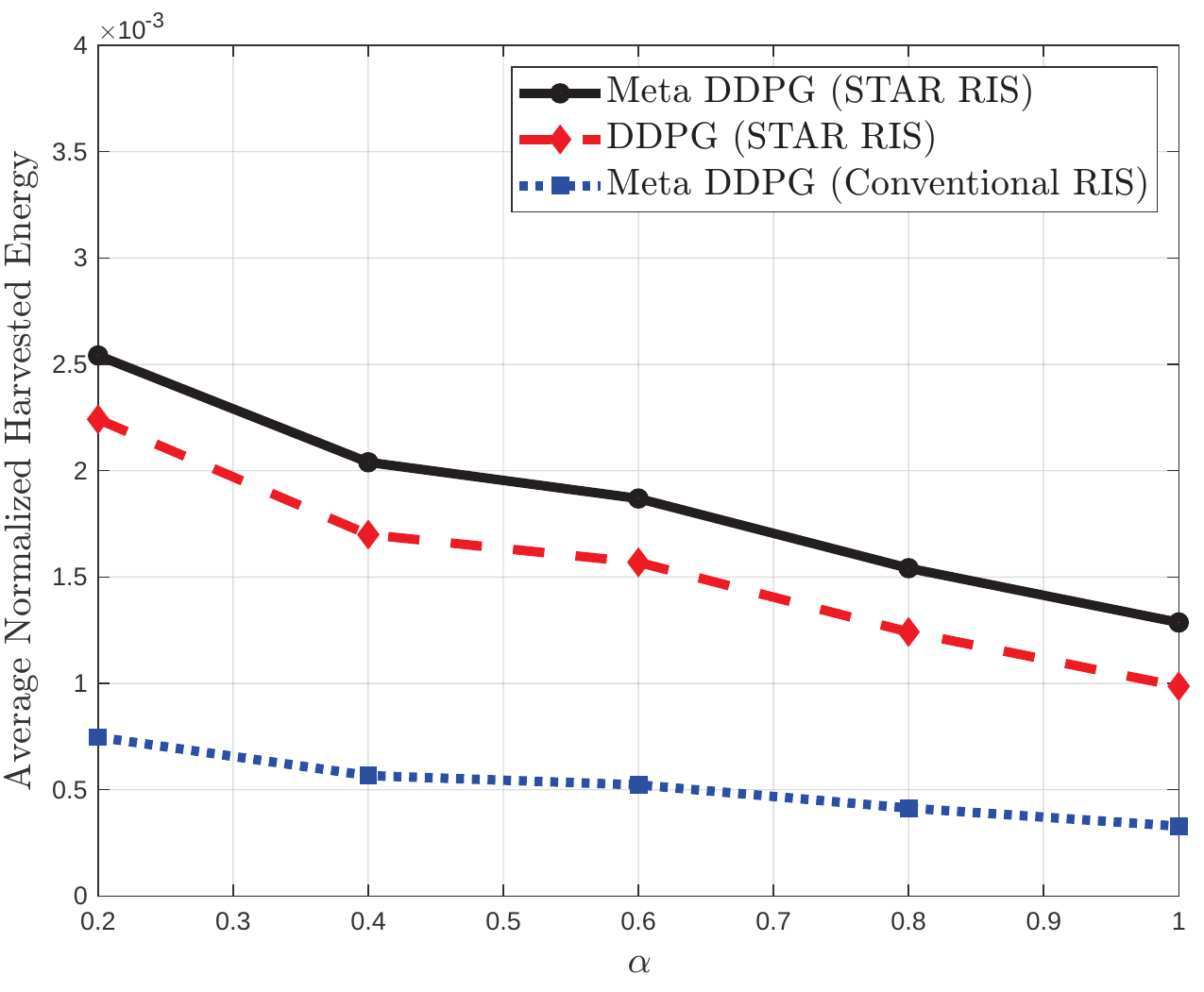}
	\caption{\footnotesize{The ANHE versus $\alpha$.}}
	\label{AAHE}
\end{figure}

				In Fig. \ref{EHADR}, the tradeoff between the data rate and average normalized harvested energy (ANHE) for various scenarios is investigated. As can be observed, the data rate decreases with increasing ANHE. This result validates the conflict between the goal of maximizing ANHE and maximizing the data rate. It is noteworthy that by integrating the Meta-DDPG approach with STAR-RIS, a more favorable trade-off condition appears.
				When combined with STAR-RIS, the Meta-DDPG approach consistently achieves the highest data rates compared to the Meta-DDPG approach coupled with conventional RIS. This observation indicates that integrating STAR-RIS with the Meta-DDPG algorithm significantly enhances data rate performance. Indeed, it is significant to point out that data rate obtained when the DDPG approach is combined with STAR-RIS is lower than the data rate achieved by the Meta-DDPG approach in conjunction with STAR-RIS. This observation implies that the utilization of the Meta-learning variant of DDPG, in combination with STAR-RIS, yields superior data rate performance compared to the standard DDPG approach.
In the following, a simulation is conducted to assess the ANHE across a range of different values for $\alpha$, as depicted in Fig. \ref{AAHE}. This figure reinforces the point that the integration of the Meta-DDPG approach with STAR-RIS yields acceptable results and enhances energy harvesting performance.

\section{Complexity Analysis}
In this paper, the computational complexity of the Meta-DDPG algorithm is assessed using Big-O notation. It can be evaluated for both training and operational phases as illustrated in Table \ref{table:questions}, where $N_{t}$ and $S_{b}$ denote the number of tasks and sampled batch, respectively. Moreover, $N_{L}$ and $\delta_{i}$ represent the number of neural network layers and the number of neurons in the $i$-th layer.
\begin{table} [h!]
	\caption{The order of complexity in two modes}
	\label{table:questions}
	\centering
	\begin{tabular}{|c|p{0.33\textwidth}|}\hline
		& Complexity
		\\   \hline
		\multirow{1}{*}{Training mode}
		& $\mathcal{O}(T\times E\times N_{t}\times S_{b}\times \sum_{i=0}^{N_{L}}\delta_{i}\delta_{i+1})$ \\
		\hline
		\multirow{1}{*}{Running mode}
		&  $\mathcal{O}(T \times E\times S_{b} \times \sum_{i=0}^{N_{L}}\delta_{i}\delta_{i+1})$ \\
		\hline
	\end{tabular}
	\label{table_1}
\end{table}

\section{Conclusion}
In this paper, a novel approach was proposed to enhance the performance of wireless system via the deployment of an STAR-RIS. More specifically, an STAR-RIS SWIPT-assisted wireless MISO system with RSMA is considered, in which the Meta-DDPG
approach is applied to solve the non-convex NP-hard problem. In this setup, the users are separated into two groups, namely IDRs and EHRs which receive the information and energy from the multiple-antenna BS, respectively. Furthermore, the IDRs's sum rate maximization problem as well as the maximization of total harvested energy at the EHRs is solved by applying the Meta-DDPG algorithm. Simulation results confirmed the superiority of the proposed Meta-DDPG algorithm in both data rate and harvested energy enhancements compared to the conventional DDPG.

\bibliographystyle{IEEEtran}
{\footnotesize
\bibliography{IEEEabrv,ref}

\end{document}